\documentclass[12pt]{article}
\usepackage{amssymb, amsmath, amsfonts}
\usepackage[cp1251]{inputenc}       
\usepackage[russian, english]{babel}

\newtheorem{thm}{Theorem}
\newtheorem{pro}[thm]{Proposition}
\newtheorem{lem}[thm]{Lemma}

\font\myfont=msbm10 scaled \magstep1
\def\CC{\hbox{\myfont\char'103}}

\def\baselinestretch{1.5}
\numberwithin{equation}{section}
\begin{document}

\begin{center}
{\Large Formal diagonalization of the discrete Lax operators and
construction of conserved densities and symmetries for dynamical systems}
\end{center}

\begin{center}
{Ismagil Habibullin}\footnote{e-mail: habibullinismagil@gmail.com}\\

{Ufa Institute of Mathematics, Russian Academy of Science,\\
Chernyshevskii Str., 112, Ufa, 450077, Russia}\\
and Bashkir State University, 
32, Validi Str., 450076, Ufa, Russia,

 {Marina Yangubaeva}\footnote{e-mail: marina.yangubaeva@mail.ru}\\
Bashkir State University, 
32, Validi Str., 450076, Ufa, Russia
 
\end{center}

\begin{abstract}
An alternative method of constructing the formal diagonalization for the discrete Lax operators is proposed which can be used to calculate conservation laws and in some cases generalized symmetries for discrete dynamical systems. Discrete potential KdV equation, lattice derivative nonlinear Schr\"odinger equation, dressing chain, Toda lattice are considered as illustrative examples. For the Toda lattice on a quad graph  corresponding to the Lie algebra $A_1^{(1)}$ infinite series of conservation laws are described. Systems of quad graph equations are represented including lattice versions of the ``matrix" NLS and ``vector'' derivative NLS equations.
\end{abstract}

{\it Keywords:} Lax pair, formal asymptotics, conserved quantities, symmetries, systems of quad graph equations, lattice NLS equation\\
\def\baselinestretch{1.5}

PACS number: 02.30.Ik

\newpage
\tableofcontents

\section{Introduction}

Systems of linear ordinary differential equations with additional (spectral) parameter are well studied (see, for instance, \cite{Kodd, Vazov}). Asymptotic behaviour of such systems as the parameter approaches a point of singularity  is investigated in details. The last four decades the interest to linear systems  is renewed due to important applications in soliton theory (see monographs \cite{Abl}--\cite{{ZSh79}}). Actually the  coefficients of the formal expansion of Lax operator's eigenfunctions generate conservation laws for a class of nonlinear dynamical systems. The conventional  algorithm for computing the asymptotic expansion of a solution to the equation $Ly=(\frac{d}{dx}+q(x)-a\lambda)y=0$ was simplified in \cite{Dri} by applying the method of formal diagonalization. Here $a$ is a matrix with pairwise different eigenvalues. The idea of diagonalization of the Lax operator turned out to be very fruitful, it allowed the authors to find a deep connection between the infinite-dimensional Lie algebras of Kac-Moody and integrable systems of differential equations. Note that the method of formal diagonalization can be applied to any operator of the form 
\begin{equation}\label{do}
L=\frac{d}{dx}+q(x,\lambda)
\end{equation}
with
\begin{equation}\label{doc}
q(x,\lambda)=-a\lambda+\sum_{i=0}^{\infty}q^{(i)}\lambda^{-i}.
\end{equation}
To the best of our knowledge the problem of reducing generic operator (\ref{do}) to the canonical form (\ref{do})-(\ref{doc}) by a linear transformation is still open. However for special cases such kind of transformation can be found (see, for example, the transformation below reducing the system (\ref{vslax}) to the form  (\ref{newvslax})).
Currently discrete integrable nonlinear models attract the attention of many researchers. Lax pair is one of the main tools to investigate nonlinear integrable equations. In this context,  studying of equations of the form \begin{equation}\label{ds}
y(n+1)=F({\bf u},\lambda)y(n)
\end{equation}
is of particular relevance. Here ${\bf u}={\bf u}(n)$ is a vector-function of the discrete argument.
To our knowledge the  problem of  asymptotic behavior for (\ref{ds}) 
is still less studied even in the case when $F({\bf u},\lambda)$ is a polynomial on the complex parameter $\lambda$. Among the related results  it is worth mentioning the method of reducing the equation (\ref{ds}) to the nonlinear Riccati type discrete equation (see, for instance, \cite{Nov}, \cite{Fad}) which provides an effective tool to construct conservation laws. To the systems with defects the method is applied in \cite{Caudr, habkundu}.  Recently this approach was adopted to quad graph equations \cite{Cheng1}, \cite{Cheng2}.

Another important step was done by A.V.Mikhailov. He proposed the following scheme of
diagonalization of discrete linear equations which realize the
Darboux transformation for a linear differential equations. It is 
observed in \cite{Mikh2012} that the formal series diagonalizing the differential
operator diagonalizes its discrete counterparts as well. Actually
this scheme allows to construct conservation laws for a large class
of integrable differential-difference and partial difference equations. The scheme is successfully realized in  \cite{GMY}.

In the present article the Drinfeld-Sokolov formalism developed in \cite{Dri} is applied directly to discrete linear equations without any assumption about existence
of the differential operator connected with the discrete operator
given. It is shown that conservation laws for discrete dynamical systems and in some cases generalized symmetries  can be evaluated by using this approach. 

The article is organised as follows. In Section 2 we prove Proposition \ref{prop1}, containing the explanation of the method of diagonalization in the discrete case. In Section 3 an algorithm for calculation of conservation laws for discrete and semi-discrete dynamical systems via formal diagonalization is discussed.  Applications of the algorithm are illustrated in \S 4,5 where diagonalization is found for certain models. In section 6 the diagonalization method is used to calculate the hierarchy of generalized symmetries, associated with the operator of a particular form. In \S 7 systems of discrete equations are discussed including lattice versions of the ``matrix" NLS equation and ``vector" derivative NLS equation.

\section{Formal diagonalization  of discrete linear systems.}

The first step in our algorithm is to bring the equation (\ref{ds}) to some canonical form convenient for evaluating asymptotic expansions. We rewrite equation (\ref{ds}) as follows
\begin{equation}\label{main}
y(n+1)=P({\bf u},\lambda)Zy(n),
\end{equation}
where $Z=diag(\lambda^{\gamma_1},\lambda^{\gamma_2},\dots,\lambda^{\gamma_N})$ is a diagonal matrix where the exponents are pairwise different integers\footnote{The scheme suggested is easily generalized to the case when some of the exponents coincide, see Section 7 below.}. For the sake of simplicity the exponents are  ordered as follows 
 $\gamma_1>\gamma_2>\dots>\gamma_N$. We assume that $P=P({\bf u},\lambda)$ is a function ranging in $\CC^{N\times N}$, depending on a discrete variable $n$ and a parameter $\lambda$. 
Additionally we require that at least one of the following conditions valid: 
for any integer $n$ function $P=P({\bf u},\lambda)$ is analytical either in a neighborhood of the point $\lambda =\infty$ or in a neighborhood of the point $\lambda =0$. Some extra conditions will be listed in Proposition \ref{prop1} below. Explain briefly the essence of the canonical form and discuss how to find it for a given system. The representation (\ref{main}) is originated by the discrete version of the Zakharov-Shabat dressing transform. Let us recall briefly the scheme of the discrete dressing. For a given matrix-valued smooth function $r=r(\lambda)$ defined on a contour (for the simplicity take the circumference $|\lambda|=1$) we define the function $r(n,\lambda)=Z^nr(\lambda)Z^{-n}$. Then construct a pair of non-degenerate matrix-valued functions $\phi(n,\lambda)$, $\psi(n,\lambda)$ defined on the same contour $|\lambda|=1$ which admit for any $n$ non-degenerate analytical continuation from the contour into the domain $|\lambda|<1$ and respectively, the domain $|\lambda|>1$ and satisfy for $|\lambda|=1$ the following conjugation relation
\begin{equation}\label{rp}
\phi(n,\lambda)r(n,\lambda)=\psi(n,\lambda).
\end{equation}
If $r=r(\lambda)$ is sufficiently close to the identity matrix then the problem (\ref{rp}) can effectively be solved. To provide the uniqueness one has to put additional normalizing condition. The simplest one is of the form
\begin{equation}\label{nc}
\psi(n,\infty)=1.
\end{equation}

Now the dressed potential is easily found
\begin{equation}\label{dp}
F({\bf u},\lambda):=\phi(n+1,\lambda)Z\phi^{-1}(n,\lambda)=\psi(n+1,\lambda)Z\psi^{-1}(n,\lambda).
\end{equation}
However the problem arises to describe the manifold $M_Z$ of all possible values of the potential $F({\bf u},\lambda)$. For example, in $2\times 2$ case with $Z=diag(\lambda, 1)$ and normalization (\ref{nc}) we have
\begin{equation}\label{fexample}
F({\bf u},\lambda)=\left(\begin{array}{cc}
\lambda+uv&v\\
u&1
\end{array}
\right).
\end{equation}
Here evidently ${\bf u}=(u,v)$ is a functional parameter.

Thus the discrete dressing is a map converting a matrix $Z(\lambda)$ to a matrix $F({\bf u},\lambda)$ which serves as a potential of the equation (\ref{ds}). It is proved in \cite{Hab1985} that the potential $F({\bf u},\lambda)$ is represented as products of the form
\begin{equation}\label{dress}
F({\bf u},\lambda)=\alpha_+(u,\lambda)Z\beta_+(u,\lambda)=\alpha_-(u,\lambda)Z\beta_-(u,\lambda)
\end{equation}
where $\alpha_{\pm}$, $\beta_{\pm}$ are triangular matrices rationally depending on $\lambda$ having non-degenerate limits for $\lambda^{\pm1}\rightarrow0$.  Therefore for such a potential the discrete equation (\ref{ds}) can be brought by linear transformations $z=\beta_{\pm}y$ to two different canonical forms $z_{n+1}=D_n(\beta_{\pm})\alpha_{\pm}Zz_n$. It is convenient to work with the operator $L=D_n^{-1}F$ generated by the equation (\ref{ds}) instead of the equation itself. Here $D_n$ is the shift operator acting due to the rule $D_nf(n)=f(n+1)$. The reasonings above show that under the conjugation $L_{\pm}=\beta_{\pm}L\beta^{-1}_{\pm}$ the operator $L$ is reduced to two different canonical forms  $L_{\pm}=D^{-1}_nP_{\pm}Z$, where $P_{\pm}=D_{n}(\beta_{\pm})\alpha_{\pm}$. For the operator with the potential (\ref{fexample}) the canonical form  is studied in Section 4.1.

Proposition \ref{prop1} below demonstrates that the canonical form (\ref{main}) is very suitable for studying asymptotics of the Lax operator's eigenfunctions. Note that usually discrete Lax operators are given in a polynomial form (\ref{ds}). Thus to apply the proposition one has to convert one form to another. To solve this problem one should first find at least one of the representations (\ref{dress}). Examples below show that the transformation searched is connected with the gauge transform and subsequent triangular factorization of the potential $F({\bf u},\lambda)$.

\begin{pro}\label{prop1}. Suppose that for any integer $n$ function $P({\bf u}(n),\lambda)$ is analytical in a neighborhood of the point $\lambda=\infty$:
\begin{equation}\label{Pnl}
P({\bf u}(n),\lambda)=P_0(n)+\lambda^{-1}P_1(n)+\lambda^{-2}P_2(n)+\dots
\end{equation}
and the following regularity conditions are satisfied. All of the principal minors of the matrix $P({\bf u}(n),\infty)$ do not vanish
\begin{equation}\label{infty}
det_jP({\bf u}(n),\infty)\neq 0\quad for \quad j=1,2,\dots,N \quad and \quad n\in (-\infty,+\infty).
\end{equation}
Then there exists a formal series $T=\sum_{i\geq 0}T_i\lambda^{-i}$ such that operator $L_0:=T^{-1}LT$ is of the form $L_0=D^{-1}_nhZ$ where $h$ is a formal series with diagonal coefficients:
\begin{equation}
h=h_0+h_1\lambda^{-1}+h_2\lambda^{-2}+\dots.
\end{equation}
The series $T$ is not unique. It is defined up to multiplication by a formal series with diagonal coefficients. It can be chosen in such a way that any coefficient $T_i$ and $h_i$ depends on a finite number of the matrices $\left\{P_j({\bf u}(n+k))\right\}.$

Similarly, if function $P=P({\bf u}(n),\lambda)$ is analytical in a neighborhood of the point $\lambda =0$ 
\begin{equation}\label{Pn0}
P({\bf u}(n),\lambda)=P'_0(n)+\lambda P'_1(n)+\lambda^{2}P'_2(n)+\dots
\end{equation}
and  all of the principal minors of the matrix $P^{-1}({\bf u}(n),0)$ are different from zero, i.e.
\begin{equation}\label{zero}
det_jP^{-1}({\bf u}(n),0)\neq 0\quad for \quad j=1,2,\dots,N \quad and \quad n\in (-\infty,+\infty),
\end{equation}
then there exists a formal series $T'=\sum_{i\geq 0}T'_i\lambda^{i}$ such that operator $L'_0:=T'^{-1}LT'$ is of the form $L'_0=D^{-1}_nh'Z$ where $h'$ is a formal series with diagonal coefficients:
\begin{equation}
h'=h'_0+h'_1\lambda+h'_2\lambda^{2}+\dots.
\end{equation}
The series $T'$ can be chosen in such a way that any coefficient $T'_i$ and $h'_i$ depends on a finite number of the matrices $\left\{P'_j({\bf u}(n+k))\right\}.$
\end{pro}
{\bf Proof.}\footnote{The scheme of the proof has shortly been discussed in the article  \cite{Hab1985} by one of the authors.} Prove the first case, supposing that function $P({\bf u}(n),\lambda)$ is analytical around the point $\lambda=\infty$, the second case is proved in a similar way. The objects searched: series $T$ and operator $L_0=D^{-1}_nhZ$ satisfy the equation $LT=TL_0$ which implies $D^{-1}_nP({\bf u}(n),\lambda)Z T=TD^{-1}_nhZ$ or, after some transformation, it is led to the form
\begin{equation}\label{proof1}
D_n(T)h=PZTZ^{-1}.
\end{equation}
Hence the function $\bar T(n,\lambda):=ZT(n,\lambda)Z^{-1}$ due to the relation $\bar T(n,\lambda)=P^{-1}({\bf u}(n),\lambda)D_n(T)h$ is also analytical at infinity then we have
\begin{equation}\label{barT}
\bar T(\lambda)=\bar T_0+\bar T_1\lambda^{-1}+\bar T_2\lambda^{-2}+\dots.
\end{equation}
It follows from the formula
\begin{equation}\label{ZTZ}
ZTZ^{-1}=\left(
\begin{array}{cccc}
T_{1,1}&\lambda^{\gamma_{1,2}}T_{1,2}&\dots&\lambda^{\gamma_{1,N}}T_{1,N}\\
\lambda^{\gamma_{2,1}}T_{2,1}&T_{2,2}&\dots&\lambda^{\gamma_{2,N}}T_{2,N}\\
\dots&\dots&\dots&\dots\\
\lambda^{\gamma_{N,1}}T_{N,1}&\lambda^{\gamma_{2,N}}T_{2,N}&\dots&T_{N,N}
\end{array}
\right)
\end{equation}
where $\gamma_{i,j}=\gamma_i-\gamma_j$ that the entry $\bar T_{i,j}$ of the matrix $\bar T$ is expressed through the corresponding entry of $T$ as
\begin{equation}
\bar T_{i,j}= \lambda^{\gamma_{i,j}}T_{i,j}=\sum^{\infty}_{k=max\{0,\gamma_{i,j}\}}T_{k,i,j}\lambda^{-k+\gamma_{i,j}},
\end{equation}
therefore if $i<j$ then 
\begin{eqnarray}
\bar T_{i,j}=\sum^\infty_{k=\gamma_{i,j}} T_{k,i,j}\lambda^{-k+\gamma_{i,j}}\label{i<j}
\end{eqnarray}
since for $0\leq k<\gamma_{i,j}$ we have  $T_{k,i,j}=0$.
Therefore the matrix $T_0$ in the expression $T=\sum_{k\geq 0} T_k\lambda^{-k}$ is lower triangular. Similarly one can check that matrix $\bar T_0$ is upper triangular. Indeed, we have for $i>j\quad \bar T_{i,j}=\lambda^{-|\gamma_{i,j}|}T_{i,j}(\lambda)\to 0$ for $\lambda\to\infty$ because $\gamma_{i,j}<0$ for $i>j$.

Now by comparing the first coefficients in the equation (\ref{proof1}), rewritten as follows
\begin{equation}\label{ThPT}
D_n(T_0+\lambda^{-1}T_1+\dots)(h_0+\lambda^{-1}h_1+\dots)= (P_0+\lambda^{-1}P_1+\dots)(\bar T_0+\lambda^{-1}\bar T_1+\dots)
\end{equation}
one gets a nonlinear equation for defining matrices $T_0, h_0, \bar T_0$
\begin{equation}\label{cond}
D_n(T_0)h_0=P_0\bar T_0,
\end{equation}
which is nothing else but the problem of multiplicative Gauss decomposition for the  matrix $P_0$ as a product of three matrices: lower triangular $D_n(T_0)$, diagonal $h_0$ and upper triangular $\bar T_0^{-1}$. The problem has a solution due to regularity condition (\ref{infty}) (see, for instance, \cite{gan}). Suppose that diagonal entries of the matrix $T(0)$ are chosen as follows $diag T_0=(1,1,\dots,1)$. Then due to the relation $diag T_0=diag \bar T_0$ the problem \eqref{cond} is uniquely solved and hence matrices $T_0$, $h_0$ and $\bar T_0$ are found.

We request that all diagonal entries of the matrices $T_k$ with $k>0$ are equal to zero. Decompose any of the matrices $T_k, \bar T_k, k>0$ into a sum of lower and upper triangular matrices
\begin{equation}
T_k=T_{kL}+T_{kU},\quad \bar T_k=\bar T_{kL}+\bar T_{kU}
\end{equation}
with vanishing diagonals. Note that the matrices $T_{1U}$ and $\bar T_{1L}$ are easily found. Really, it was observed above that $T_{1,i,j}=0$ for $i<j$ and $\gamma_{i,j}>1$. Since there exists $\gamma_{i,j}=1$ for $i<j$ then $T_{1,i,j}=\bar T_{0,i,j}$ and this entry of the matrix $T_{1U}$ is found in the previous step. Similarly, since $\bar T_{i,j}=\lambda^{\gamma_{i,j}}T_{i,j}$ then $\bar T_{1,i,j}=0$ for $i>j$ and $\gamma_{i,j}<-1$. For the entries $\bar T_{1,i,j}$ with $\gamma_{i,j}=-1$ we have $\bar T_{1,i,j}=T_{0,i,j}$ therefore those entries of $\bar T_{1L}$ coincide with the entries of $T_0$ found earlier. To look for the other entries of the matrices $T_1,$ $\bar T_1$ derive the equation 
\begin{equation}
D_n(T_0)h_1+D_n(T_1)h_0=P_0\bar T_1+P_1\bar T_0
\end{equation}
which can be rewritten as follows
\begin{equation}
h_1h^{-1}_0+D_n(T^{-1}_0T_{1L})-h_0\bar T^{-1}_0\bar T_{1U}h^{-1}_0=H_1
\end{equation}
where $H_1=D_n(T^{-1}_0)P_1\bar T_0h^{-1}_0-D_n(T^{-1}_0T_{1U})+h_0\bar T^{-1}_0\bar T_{1L}h^{-1}_0$ contains either already found or given matrices.

In order to find the unknown matrices $h_1, DT_{1L}$ and $\bar T_{1U}$ one should first decompose the matrix $H_1$ into a sum of three matrices: a diagonal matrix which equals $h_1h^{-1}_0$, a lower triangular nilpotent matrix $D_n(T^{-1}_0T_{1L})$ and an upper diagonal nilpotent matrix $h_0\bar T^{-1}_0\bar T_{1U}h^{-1}_0$. Iterating this procedure one can find all of the coefficients of the series $T$ and $h$. Indeed the coefficients $T_k,h_k$ are found by solving the equation
\begin{equation}
h_kh^{-1}_0+D_n(T^{-1}_0T_{kL})-h_0\bar T^{-1}_0\bar T_{kU}h^{-1}_0=H_k,
\end{equation}
where the term $H_k$ contains matrices found in the previous steps. To determine the unknowns $T_{kL}$, $\bar T_{kU}$, $h_k$ it is enough to decompose $H_k$ as a sum of lower triangular, diagonal and upper diagonal matrices, i.e. one has to solve the problem of linear Gauss decomposition. 

Suppose that the operator $\tilde L_0=\tilde T^{-1} L\tilde T$ is of the form $\tilde L_0=D^{-1}\tilde h Z$, and its coefficients are also diagonal matrices. Let us set $S=T^{-1}\tilde T $ and then get the equation
$S\tilde L_0 = L_0S$ which shows that $S$ is a diagonal matrix. This proves that the conjugation matrix $T$ is defined up to multiplying from the right side by a diagonal factor. 

The Proposition is proved.

\section{Conservation laws for discrete dynamical systems.}

Suppose that a dynamical system of the form
\begin{equation}\label{eps}
\epsilon(D_mD_n{\bf u},D_m{\bf u},D_n{\bf u}, {\bf u})=0
\end{equation}
where the field variable ${\bf u} (n,m)=(u_1(n,m),u_2(n,m),\dots,u_N(n,m))$ depends on two integers $n,m$, admits the Lax representation. Or in other words equation \eqref{eps} is the consistency condition of a pair of the linear equations
\begin{equation}\label{pairY}
y(n+1,m)=P([{\bf u}],\lambda)Zy(n,m), \quad y(n,m+1)=R([{\bf u}],\lambda)y(n,m).
\end{equation}
Note that the first equation is written in the canonical form. Here notation $[{\bf u}]$ means that $P,R$ depend on the variable $\bf u$ and a finite number of its shifts: $D^k_n{\bf u},$ $D^k_m{\bf u}$, where $D_m$ is the shift operator with respect to the variable $m$: $D_mf(n,m)=f(n,m+1)$. Evidently, the compatibility condition of the equations \eqref{pairY} can be rewritten as
\begin{equation}\label{condRP}
D_m(P)ZR=D_n(R)PZ.
\end{equation}
Introduce the operators $L=D^{-1}_nPZ$, $M=D^{-1}_mR$. Then equation \eqref{condRP} takes the form:
\begin{equation}\label{LM}
[L,M]=0.
\end{equation}  
Assume that matrix $P([{\bf u}],\lambda)$ satisfies the settings of Proposition \ref{prop1}, i.e. it is analytical around the point $\lambda=\infty$ for all values of ${\bf u}$ in a domain and the principal minors of the matrix $P([{\bf u}], \infty)$ differ from zero in this domain. Suppose that $R([{\bf u}],\lambda)$ is meromorphic at the vicinity of the point $\lambda=\infty$ when $\bf u$ ranges the corresponding domain. Then the operators $L,M$ are diagonalized simultaneously. Now following the scheme, proposed in \cite{Dri} one can find generating functions for conservation laws, see also \cite{Mikh2012}.  Their diagonal forms allow one to find generating functions for conservation laws. Due to Proposition \ref{prop1} we have $L_0=T^{-1}LT$ where $L_0=D^{-1}hZ$ is a diagonal operator. Denote $M_0:=T^{-1}MT$. According to our assumption above we have $M_0=D^{-1}_mS$, where $S=\sum_{i=k}^\infty S_i\lambda^i$ is meromorphic. Therefore equation $[L_0,M_0]=0$ implies
\begin{equation}\label{Sh}
D_n(S)h=D_m(h)ZSZ^{-1}.
\end{equation}
Equation \eqref{Sh} looks very similar to equation \eqref{proof1} with $S$ instead of $T$ and $D_m(h)$ instead of $P$. Since the matrix $D_m(h)$ is diagonal then all of the matrices $S_k,S_{k-1},\dots$ are also diagonal. This means that relation \eqref{Sh} generates conservation laws for the dynamical system \eqref{eps}. Indeed, we have 
\begin{equation}\label{cond_integral}
\frac{D_m(h)}{h}=\frac{D_n(S)}{S}, 
\end{equation}
or equivalently
\begin{equation}\label{log_cond_integral}
(D_m-1)\ln h=(D_n-1)\ln S. 
\end{equation}

Now by using formal expansions  
\begin{gather*}
h=h_0+h_1\lambda^{-1}+h_2\lambda^{-2}+..., \quad
S=S_0+S_1\lambda^{-1}+S_2\lambda^{-2}+...
\end{gather*}
one finds 
\begin{gather}\label{logSH}
(D_n-1)\left(\ln(S_0)+\frac{S_1}{S_0}\lambda^{-1}+\left(\frac{S_2}{S_0}-\frac{1}{2}\left(\frac{S_1}{S_0} \right)^2 \right)\lambda^{-2}+\dots\right)=\\=
(D_m-1)\left(\ln(h_0)+\frac{h_1}{h_0}\lambda^{-1}+\left(\frac{h_2}{h_0}-\frac{1}{2}\left(\frac{h_1}{h_0} \right)^2 \right)\lambda^{-2}+\dots\right).
\end{gather}

Consider a system of semi-discrete equations 
\begin{equation}\label{smds}
E([{\bf u}_t],[{\bf u}])=0
\end{equation}
relating the vector valued variable $\bf u$, its derivatives w.r.t. $t$, its shifts and shifts of the derivatives. Suppose that equation (\ref{smds}) is the consistency condition for an overdetermined system of linear equations
\begin{equation}\label{sml}
y(n+1,t)=P([{\bf u}],\lambda)Zy(n,t), \quad y_t(n,t)=A([{\bf u}],[{\bf u}_t], \lambda)y(n,t).
\end{equation}
System (\ref{sml}) implies
\begin{equation}\label{standard}
P_tZ=D_n(A)PZ-PZA
\end{equation}
or the same $[D_t-A,L]=0$ where $L=D_n^{-1}PZ$ and $D_t$ is the operator of total derivative w.r.t. $t$. Suppose that the potential $P$ satisfies the conditions of Proposition \ref{prop1}. By applying the conjugation operator acting as follows $x\rightarrow T^{-1}xT$ to the last equation one gets
\begin{equation}\label{comm}
[D_t-A_0,L_0]=0,
\end{equation}
where $A_0=-T^{-1}T_t+T^{-1}AT$ is a formal power series with coefficients being diagonal matrices, operator $L_0=D_n^{-1}hZ=T^{-1}LT$ is defined above in Proposition \ref{prop1}. Equation (\ref{comm}) determines generating functions of conserved densities for the dynamical system (\ref{smds}):
\begin{equation}\label{consdens}
D_t\ln h=(D_n-1)A_0.
\end{equation}

\section{Examples of fully discrete models}

In the examples below we show that the canonical form (\ref{main}) is effectively found by using the factorization (\ref{dress}).

\subsection{Lattice derivative Nonlinear Schr{\"o}dinger equation}

As one of the illustrative examples of application of diagonalization method above we consider the system 
\begin{equation}\label{lnls}
u_{0,1}-u_{1,0}+u_{1,1}((u_{0,1}-u_{1,0})v_{1,0}+\varepsilon)=0, \quad v_{1,0}-v_{0,1}+v((v_{1,0}-v_{0,1})u_{0,1}+\varepsilon)=0
\end{equation}
which looks very similar to that studied in \cite{Mikh}, \cite{Cheng2}. It will be shown in \S7 that (\ref{lnls}) is a discretization of the derivative Nonlinear Schr{\"o}dinger equation.  System of equations (\ref{lnls}) is the consistency condition of the following linear equations
\begin{equation}\label{linearequations}
y_{1,0}=fy, \quad y_{0,1}=gy 
\end{equation}
or the commutativity condition of the operators
\begin{equation}\label{operators}
L=D^{-1}_nf \quad \mbox{and}\quad M=D^{-1}_mg
\end{equation}
where 
\begin{equation}\label{fgidlns}
f=\left(\begin{array}{cc}
\lambda+uv&v\\
u&1
\end{array}
\right),\quad
g=\left(\begin{array}{cc}
\lambda+\varepsilon+ u_{-1,1}v&v\\
u_{-1,1}&1
\end{array}
\right)
\end{equation}
and the notations are accepted $y_{1,0}=D_ny$, $y_{0,1}=D_my$, $u_{-1,1}=D_n^{-1}D_mu$ and so on.
Operators of this form have been studied earlier in \cite{Shabat78}. It is remarkable that potential $f$ admits two different kind triangular representations allowing to get two canonical forms. Indeed,
\begin{equation}\label{+-}
f=PZ
\end{equation} 
where
\begin{equation}
P=\left(\begin{array}{cc}
1+uv\lambda^{-1}&v\\ 
u\lambda^{-1}&1
\end{array} \nonumber
\right),\quad
Z=\left(\begin{array}{cc}
\lambda&0\\
0&1
\end{array}
\right). \nonumber
\end{equation}

Since the operator $L'=D^{-1}_nPZ$ is of the canonical form we can apply to it the method of formal diagonalization from Proposition \ref{prop1}. To this end we have to look for the formal series $T=T_0+\lambda^{-1}T_1+\lambda^{-2}T_2+\dots $ and $h=h_0+\lambda^{-1}h_1+\lambda^{-2}h_2+\dots $ such that the operator $L'_0:=T^{-1}L'T$ is diagonal $L'_0=D^{-1}hZ$. Due to the general scheme to find such $T$ and $h$ one has to solve the following equation
\begin{equation}\label{mainequation}
D_n(T)h=P\bar T
\end{equation}
where $\bar T=ZTZ^{-1}=\bar T_0+\lambda^{-1}\bar T_1+\dots$.

By comparing the coefficients before the powers of $\lambda$ we find a sequence of equations 
\begin{eqnarray} 
&& D_n(T_0)h_0=P_0\bar T_0, \quad D_n(T_1)h_0+D_n(T_0)h_1-P_0\bar T_1=P_1\bar T_0,  \nonumber\\
&& D_n(T_k)h_0+D_n(T_0)h_k-P_0\bar T_k=P_1\bar T_{k-1}-\sum^{k-1}_{j=1}D_n(T_{k-j})h_j,\quad \mbox{for}\quad k\geq 2\nonumber. 
\end{eqnarray}

Here the first equation is solved just by applying the formula of Gauss decomposition. From the equation with number $k$ one can find the lower triangular part of the matrix $D_n(T_k)$, upper triangular part of $\bar T_k$ and the diagonal matrix $h_k$. The other objects in the equation are already found from the previous equations (from the equations with numbers less than $k$). As a result we find
\begin{eqnarray}
T=\left(\begin{array}{cc}
1&0\\ 
0&1
\end{array}\right)+
\left(\begin{array}{cc}
0&-v\\ 
u_{-1,0}&0
\end{array}\right)\lambda^{-1}+
\left(\begin{array}{cc}
0&uv^2-v_{1,0}\\ 
u_{-2,0}-u_{-1,0}^2v_{-1,0}&0
\end{array}\right)\lambda^{-2}+ \dots\label{Tseries}\\
h=\left(\begin{array}{cc}
1&0\\
0&1
\end{array}\right)
+\left( \begin{array}{cc}
uv&0\\ 
0&-uv
\end{array} \right)\lambda^{-1}+
\left( \begin{array}{cc}
u_{-1,0}v&0\\ 
0&u^2v^2-uv_{1,0}
\end{array} \right)\lambda^{-2}+\dots. 
\end{eqnarray}
The second operator is diagonalized as follows
\begin{gather}
S:=D_m(T^{-1})gT=
\left(\begin{array}{cc}
1&0\\
0&0
\end{array}\right)\lambda
+\left( \begin{array}{cc}
\varepsilon+u_{-1,1}v&0\\ 
0&1
\end{array} \right)+
\left( \begin{array}{cc}
u_{-1,0}v&0\\ 
0&-u_{-1,1}v
\end{array} \right)\lambda^{-1}+\nonumber\\
+\left( \begin{array}{cc}
u_{-2,0}v-u_{-1,0}^2vv_{-1,0}&0\\ 
0&u_{-1,1}^2v^2+u_{-1,1}v\varepsilon-u_{-1,1}v_{0,1}
\end{array} \right)\lambda^{-2}+\dots
\end{gather}
Let us give the first three conservation laws from the infinite sequence generated by the diagonalization, see also \cite{Cheng2}
\begin{eqnarray}
&&(D_n-1)u_{-1,1}v=(D_m-1)uv,\nonumber \\
&&(D_n-1)\left(u_{-1,0}v-\frac{1}{2}(\varepsilon+u_{-1,1}v)^2\right)=(D_m-1)\left(u_{-1,0}v-\frac{1}{2}u^2v^2\right),\nonumber \\
&&(D_n-1)\left(\frac{1}{2}u_{0,1}^2v^2_{1,0}+u_{0,1}v_{1,0}\varepsilon-u_{0,1}v_{1,1}\right)=
(D_m-1)\left(\frac{1}{2}u_{1,0}^2v_{1,0}^2-u_{1,0}v_{2,0}\right). \nonumber
\end{eqnarray}

\subsection{Generalized discrete Toda lattice corresponding to the Lie algebra $A^{(1)}_1$}

Consider the system corresponding to  the affine Lie algebra $A^{(1)}_1$ proposed recently in \cite{GHY} 
\begin{gather}\label{dtoda} 
uu_{1,1}-u_{1,0}u_{0,1}=v_{0,1}^2,\\
vv_{1,1}-v_{1,0}v_{0,1}=u_{1,0}^2.\nonumber
\end{gather}
System (\ref{dtoda}) is proved to be an integrable discretization of the Toda lattice related to the algebra $A^{(1)}_1$
\begin{gather}\nonumber 
r_{x,y}=e^{-2r+2s},\\
s_{x,y}=e^{2r-2s}.\nonumber
\end{gather}
The continuum limit is obtained in the standard way. First one has to change in (\ref{dtoda}) $u\rightarrow e^{R}$, $v\rightarrow e^{S}$. Then take $R(n,m)=r(x,y)$, $S(n,m)=s(x,y)$, $x=n\varepsilon$, $y=m\varepsilon$ and evaluate in the equation obtained the limit for $\varepsilon\rightarrow0$.

System (\ref{dtoda}) is the compatibility conditions for the equations 
\begin{equation}\label{Psi}
\Psi_{1,0}=f\Psi,\quad \Psi_{0,1}=g\Psi,
\end{equation}
where 
\[
f=\left(
\begin{array}{cc}
\frac{uv_{1,0}}{u_{1,0}v}+\lambda&-\frac{u_{1,0}v_{-1,0}}{uv}\\
-\lambda&\frac{u_{1,0}v_{-1,0}}{uv}
\end{array}\right),\quad 
g=\left(
\begin{array}{cc}
1+\frac{1}{\lambda}\frac{uv_{0,1}}{u_{0,1}v}&
\frac{1}{\lambda}\frac{v_{0,1}^2v_{-1,0}}{uu_{0,1}v}\\
\frac{u^2}{vv_{-1,1}}&\frac{v_{0,1}v_{-1,0}}{vv_{-1,1}}
\end{array}\right).
\]

For the sake of convenience put $a=\frac{v_{1,0}u}{vu_{1,0}}$, $b=\frac{u_{1,0}v_{-1,0}}{uv}$, $c=\frac{v_{0,1}u}{u_{0,1}v}$, $d=\frac{(v_{0,1})^2v_{-1,0}}{uu_{0,1}v}$, $e=\frac{u^2}{vv_{-1,1}}=\frac{v_{0,1}v_{-1,0}}{vv_{-1,1}}-1$. Then it is straightforward to check that $c(1+e)=ed$, therefore  $\det g=1+e$. Thus we have
\begin{equation}
f=\left(
\begin{array}{cc}
a+\lambda&-b\\
-\lambda&b
\end{array}\right),\quad 
g=\left(
\begin{array}{cc}
1+\frac{1}{\lambda}c&
\frac{1}{\lambda}d\\
e&1+e
\end{array}\right).
\end{equation} 

Split down the matrix $f$ into a product of three factors 

\begin{equation}
f=\alpha Z\gamma,
\end{equation}
where 
\begin{equation}
\alpha=\left(
\begin{array}{cc}
1+a\lambda^{-1}&0\\
-1&\frac{a}{1+a\lambda^{-1}}
\end{array}\right),
Z=\left(
\begin{array}{cc}
\lambda&0\\
0&\lambda^{-1}
\end{array}\right),
\gamma=\left(
\begin{array}{cc}
1&-\frac{b\lambda^{-1}}{1+a\lambda^{-1}}\\
0&b
\end{array}\right).
\end{equation}

By changing the variables as $\phi=\gamma \Psi$ reduce the first equation in (\ref{Psi}) to its canonical form 
\begin{equation}\label{tphi}
\phi_{1,0}=P(\lambda)Z\phi
\end{equation}
where
\begin{gather*}
P(\lambda)=\gamma_{10}\alpha=\left(
\begin{array}{cc}
1+a\lambda^{-1}+\frac{b_{1,0}\lambda^{-1}}{1+a_{1,0}\lambda^{-1}}&-\frac{ab_{1,0}\lambda^{-1}}{(1+a_{1,0}\lambda^{-1})(1+a\lambda^{-1})}\\
-b_{1,0}&\frac{ab_{1,0}}{1+a\lambda^{-1}}
\end{array}\right)
=\left(
\begin{array}{cc}
1&0\\
-b_{1,0}&ab_{1,0}
\end{array}\right)+\\
+\left(
\begin{array}{cc}
a+b_{1,0}&-ab_{1,0}\\
0&-a^2b_{1,0}
\end{array}\right)\lambda^{-1}+
\left(
\begin{array}{cc}
-a_{1,0}b_{1,0}&-a^3b_{1,0}-a^2a_{1,0}b_{1,0}-aa_{1,0}^2b_{1,0}\\
0&a^3b_{1,0}
\end{array}\right)\lambda^{-2}+\dots .
\end{gather*}
Evidently the factor $P(\lambda)$ is analytical at $\lambda=\infty$ and $det_jP(\infty)\neq 0$. For this system we have 
\begin{gather}\nonumber
T=\left( \begin{array}{cc}
1&0\\
-b&1
\end{array}\right)+
\left( \begin{array}{cc}
0&0\\
a_{-1,0}b+b^2&0
\end{array}\right)\lambda^{-1}+
\left( \begin{array}{cc}
0&0\\
-a_{-1,0}b_{-1,0}b-ab^2-b(a_{-1,0}+b)^2&0
\end{array}\right)\lambda^{-2}+\dots,\\\nonumber
h=\left( \begin{array}{cc}
1&0\\
0&ab_{1,0}
\end{array}\right)+
\left( \begin{array}{cc}
a+b_{1,0}&0\\
0&-ab_{1,0}(a+b_{1,0})
\end{array}\right)\lambda^{-1}+\\\nonumber
 +\left( \begin{array}{cc}
-a_{1,0}b_{1,0}&0\\
0&ab_{1,0}(a^2+a_{1,0}b_{1,0}+ab_{1,0})
\end{array}\right)\lambda^{-2}+\dots.
\end{gather}
To diagonalize the second operator we have to conjugate as follows $M_0=T^{-1}\gamma M\gamma^{-1} T$ and find the operator  $M_0=D_m^{-1}S$ where $S=D_m(T^{-1}\gamma)g\gamma^{-1}T$ or
\begin{eqnarray}\nonumber
S=\left( \begin{array}{cc}
1&0\\
0&\frac{uu_{1,1}}{u_{1,0}u_{0,1}}
\end{array}\right)+
\left( \begin{array}{cc}
-\frac{uu_{1,0}}{vv_{0,1}}&0\\
0&\frac{u^2u_{1,1}}{u_{0,1}vv_{0,1}}
\end{array}\right)\lambda^{-1}+
\left( \begin{array}{cc}
-\frac{u^2v_{1,1}}{vv_{0,1}^2}&0\\
0&\frac{u^3u_{1,1}v_{1,0}}{u_{1,0}u_{0,1}v^2v_{0,1}}
\end{array}\right)\lambda^{-2}+\dots .
\end{eqnarray}
For the functions $S,h$ found the  equation $\frac{D_n(S_0)}{S_0}=\frac{D_m(h_0)}{h_0}$ generates immediately the following conservation law
\begin{equation}\nonumber
(D_n-1)\ln\frac{uu_{1,1}}{u_{1,0}u_{0,1}}=(D_m-1)\ln\frac{uu_{2,0}}{u_{1,0}^2}.
\end{equation}
Below we give three more conservation laws for the system (\ref{dtoda})
\begin{eqnarray}\nonumber
&&(D_n-1)\left(-\frac{uu_{1,0}}{vv_{0,1}}\right)=(D_m-1)\left(\frac{uv_{1,0}}{vu_{1,0}}+\frac{vu_{2,0}}{u_{1,0}v_{1,0}}\right),\\
&&(D_n-1)\left(\frac{u^2v_{1,1}}{vv_{0,1}^2}-\frac{1}{2}\left(\frac{uu_{1,0}}{vv_{0,1}}\right)^2\right)=
(D_m-1)\left(\frac{vv_{2,0}}{v_{1,0}^2}-\frac{1}{2}\left(\frac{uv_{1,0}}{u_{1,0}v}+\frac{u_{2,0}v}{u_{1,0}v_{1,0}}\right)^2\right),\\\nonumber
&&(D_n-1)\left(-\frac{u^2v_{1,0}}{v^2v_{0,1}}-\frac{1}{2}\left(\frac{uu_{1,0}}{vv_{0,1}}\right)^2\right)=
(D_m-1)\left(\frac{vu_{2,0}}{u_{1,0}v_{1,0}}+\frac{1}{2}\left(\frac{uv_{1,0}}{u_{1,0}v}+\frac{u_{2,0}v}{u_{1,0}v_{1,0}}\right)^2\right).
\end{eqnarray}

The potentials $f,g$ in \eqref{Psi} have two points of singularity $\lambda=0,$ $\lambda=\infty$. Therefore system \eqref{dtoda} admits two series of conservation laws. In order to find the second series we have to deal with the second equation in \eqref{dtoda}. The ooperator $M=D_m^{-1}g$ can be reduced to the canonical form by using the following representation 
\begin{equation}\label{canonGtoda}
g(\lambda)=\left(\begin{array}{cc}
c+\xi^{-1}&0\\e\xi^{-1}&1+e
\end{array} \right)
Z(\xi)
\left(\begin{array}{cc}
1&\frac{d}{c+\xi^{-1}}\\0&\frac{1}{c+\xi^{-1}}
\end{array} \right),
\end{equation}
where $\xi=\lambda^{-1}$, $Z(\xi)=diag(\xi,\xi^{-1})$.

By applying the diagonalization to $M$ one gets another series of conservation laws. Give the first three of them 
\begin{eqnarray}
&&(D_n-1)\ln\frac{uv_{1,0}}{u_{1,0}v}=(D_m-1)\ln\frac{uv_{0,1}}{u_{0,1}v},\nonumber\\
&&(D_n-1)\left(\frac{uv_{0,2}}{u_{0,1}v_{0,1}}+\frac{u_{0,1}v}{uv_{1,0}}\right)=
(D_m-1)\left(\frac{u_{1,0}v}{uv_{1,0}}-\frac{v^2v_{1,1}}{uu_{1,0}v_{1,0}}\right),\nonumber\\
&&(D_n-1)\ln\frac{u_{0,2}v_{-1,0}v_{0,1}^2}{u_{0,1}vv_{-1,1}v_{0,2}}=
(D_m-1)\ln\frac{u_{1,1}v_{1,0}v_{0,1}v_{-1,0}}{u_{0,1}v^2v_{1,1}}.\nonumber
\end{eqnarray}

\subsection{Discrete potential Korteweg-de Vries equation}

Concentrate on the discrete potential Korteweg-de Vries equation \cite{esta}
\begin{equation}\label{dpKdV}
(u_{1,1}-u)(u_{1,0}-u_{0,1})=4c^2.
\end{equation} 
This equation is known to be integrable, it admits the Lax pair. Infinite series of conserved quantities for it were described via the Gardner method in \cite{Ras}, hierarchy of higher symmetries constructed in \cite{mwx}. Lax pair for (\ref{dpKdV}) is given by the linear discrete equations \cite{nij}
\begin{equation}\label{psi10}
\Psi_{1,0}=f\Psi,
\quad
\Psi_{0,1}=g\Psi,
\end{equation}
where the potentials $f,g$ polynomially depend on the spectral parameter. They are matrices of the form
\begin{equation}
f=\left(
\begin{array}{cc}
-u_{1,0}&1\\
-\lambda^{-2}-uu_{1,0}&u
\end{array}
\right),\quad
g=\left(
\begin{array}{cc}
-u_{0,1}&1\\
\lambda^{-2}+4c^2-uu_{0,1}&u
\end{array}
\right). 
\end{equation}
By setting $\Psi=\lambda^{-n}\psi$ one gets $\psi_{1,0}=F\psi$ where potential 
\begin{equation}
F=\left( \begin{array}{cc}
-u_{1,0}\lambda &\lambda\\
-\lambda^{-1}-uu_{1,0}\lambda&u\lambda
\end{array}\right)
\end{equation}
is a matrix with the determinant equal to the unity.
Now our main problem is to bring the first linear equation of (\ref{psi10}) to the canonical form (\ref{main}). To this end we decompose the potential $F$ into a product of triangular and diagonal matrices
$F=\alpha Z\gamma$, where
\begin{equation}
\alpha=\left(\begin{array}{cc}
1&0\\
u+\frac{1}{u_{1,0}}\lambda^{-2}&-\frac{1}{u_{1,0}}
\end{array}\right),\quad
Z=\left(\begin{array}{cc}
\lambda&0\\
0&\lambda^{-1}
\end{array}\right), \quad
\gamma=\left(\begin{array}{cc}
-u_{1,0}&1\\
0&1
\end{array}\right).
\end{equation}
Changing as follows $y=\gamma\psi$ we convert the equation to the desired canonical form 
\begin{equation}\label{dpkdvcan}
y_{1,0}=P(\lambda)Zy
\end{equation}
where $P(\lambda)=\gamma_{1,0}\alpha$ or in the coordinate form
\begin{gather*}
P(\lambda)=\left( 
\begin{array}{cc}
u-u_{2,0}+\frac{1}{u_{1,0}}\lambda^{-2}&-\frac{1}{u_{1,0}}\\
u+\frac{1}{u_{1,0}}\lambda^{-2}&-\frac{1}{u_{1,0}}
\end{array}
\right)
=\left( 
\begin{array}{cc}
u-u_{2,0}&-\frac{1}{u_{1,0}}\\
u&-\frac{1}{u_{1,0}}
\end{array}
\right)+
\left( 
\begin{array}{cc}
\frac{1}{u_{1,0}}&0\\
\frac{1}{u_{1,0}}&0
\end{array} \right) \lambda^{-2}.
\end{gather*}

In virtue of the Proposition \ref{prop1} we can diagonalize the operator $L=D_n^{-1}P(\lambda)Z$ for $det_1P(\lambda)=\frac{u_{2,0}}{u_{1,0}}\neq0$, i.e. find the formal series $T$ and $h$ 
\begin{eqnarray}\label{T0_KdV}
T=\left(\begin{array}{cc}
1&0\\\frac{u_{-1,0}}{u_{-1,0}-u_{1,0}}&1
\end{array}\right)+\left(\begin{array}{cc}
0&\frac{1}{u_{1,0}(u-u_{2,0})}\\
\frac{u_{1,0}}{(u_{-1,0}-u_{1,0})^2(u_{-2,0}-u)}&0
\end{array}\right)\lambda^{-2}+\dots,\\
 h=\left(\begin{array}{cc}
u-u_{2,0}&0\\0&\frac{u_{2,0}}{u_{1,0}(u-u_{2,0})}
\end{array}\right)+
\left(\begin{array}{cc}
-\frac{1}{u_{-1,0}-u_{1,0}}&0\\
0&\frac{uu_{1,0}-u_{1,0}u_{2,0}+u_{2,0}u_{3,0}}{u_{1,0}^2(u-u_{2,0})^2(u_{1,0}-u_{3,0})}
\end{array}\right)\lambda^{-2}+\dots
\end{eqnarray}
such that $L=TL_0T^{-1}$, $L_0=D_n^{-1}hZ$. The second equation of the Lax pair (\ref{psi10}) under the same sequence of linear transformations turns into
the equation $y_{0,1}=\gamma_{0,1}g\gamma y$. Therefore dpkdv equation (\ref{dpKdV}) is equivalent to the commutativity condition of the operators $L$, $M$, where $M=D^{-1}_m\gamma_{0,1}g\gamma$. Let us denote $M_0=T^{-1}MT$. It is easy to see that $M_0$ is an operator of the form $M_0=D^{-1}_m S$, where $S$ is a formal series evaluated due to the formula $S=D_m(T^{-1})\gamma_{0,1}g\gamma^{-1}T$. Let us give the first terms  of $S$ in an explicit form
\begin{gather*}
S=\left(\begin{array}{cc}
u-u_{1,1}&0\\
0&\frac{u_{1,1}(u_{1,0}-u_{0,1})}{u_{1,0}}
\end{array}\right)+\\
+\left(\begin{array}{cc}
\frac{1}{u_{-10}-u_{10}}&0\\
0&\frac{u_{-11}(u-u_{-11})(u_{-10}-u_{10})}{u_{10}(u_{-11}-u_{11})^2(u_{01}-u_{21})}-
\frac{u_{11}}{u_{10}(u_{-11}-u_{11})}-\frac{u_{11}u_{-10}(u-u_{-11})}{u_{10}^2(u_{-11}-u_{11})(u-u_{20})}
\end{array}\right)\lambda^{-2}+\dots.
\end{gather*}

The commutativity relation of the operators $L_0$ and $M_0$ yields $D_mhS=D_nS h$. By taking the logarithm one reduces the latter to the form convenient to evaluate conservation laws $(D_n-1)\ln S=(D_m-1)\ln h$. Comparing coefficients before the powers of $\lambda$ one gets an infinite sequence of the conservation laws. Let us show the first three of them (see, also \cite{Ras})
\begin{eqnarray}
&&(D_n-1)\ln(u-u_{11})=(D_m-1)\ln(u-u_{20}), \nonumber \\
&&(D_n-1)\ln\frac{u_{11}(u-u_{-11})(u_{-10}-u_{10})}{u_{10}(u_{-11}-u_{11})}=(D_m-1)\ln\frac{u_{20}}{u_{10}(u-u_{20})}, \nonumber \\
&&(D_n-1)\left(\frac{1}{(u_{-10}-u_{10})(u-u_{11})}\right)=(D_m-1)\left(-\frac{1}{(u_{-10}-u_{10})(u-u_{20})}\right). \nonumber 
\end {eqnarray}

Note that series $T, h, S$ are found in a different way by A.V. Mikhailov in \cite{Mikh2012} by using an additional differential operator connected with (\ref{psi10}).

\subsection{Levi-Yamilov dressing chain}

Consider the discrete equation of the form \cite{SY2009}
\begin{equation}\label{disDres}
(u_{1,0}-\alpha)(u-\alpha)=(u_{1,1}-\alpha_{0,1})(u_{0,1}-\alpha_{0,1}).
\end{equation}
Equation \eqref{disDres} admits the Lax pair consisting of two linear discrete equation
\begin{equation}\label{LYLax}
\Psi_{1,0}=f\Psi,\quad \Psi_{0,1}=g\Psi,
\end{equation}
where
\begin{equation}
f=\left(\begin{array}{cc}
\lambda &-v\\
1&0
\end{array}\right),\quad
g=\frac{1}{u_{0,1}-\alpha_{0,1}}\left(\begin{array}{cc}
\lambda(u_{0,1}-\alpha_{0,1})&2\alpha_{0,1}(u_{0,1}^2-\alpha^2_{0,1})\\
-2\alpha_{0,1}&\lambda(u_{0,1}+\alpha_{0,1})
\end{array}\right),
\end{equation}
\begin{equation*}
v=(u_{1,0}-\alpha)(u-\alpha)=(u_{1,1}-\alpha_{0,1})(u_{0,1}-\alpha_{0,1}).
\end{equation*}
Split down the matrix $f$ into a product of three factors 
\begin{equation*}
f=\alpha Z\beta,
\end{equation*}
where
\begin{equation*}
\alpha=\left(\begin{array}{cc}
1 &0\\
\lambda^{-1}&v
\end{array}\right),\quad 
\beta=\left(\begin{array}{cc}
1 &v\lambda^{-1}\\
0&1
\end{array}\right),\quad 
Z=\left(\begin{array}{cc}
\lambda &0\\
0&\lambda^{-1}
\end{array}\right).
\end{equation*}
The first equation in (\ref{LYLax}) is reduced to the canononical form $\psi_{1,0}=P(\lambda)Z\psi$ by the transformation $\psi=\beta\Psi$ where
\begin{equation*}
P=\beta_{1,0}\alpha=
\left(\begin{array}{cc}
1-v_{1,0}\lambda^{-2} &-vv_{1,0}\lambda^{-2}\\
\lambda^{-1}&v
\end{array}\right).
\end{equation*}
Matrix $P$ satisfies the requirements of the Proposition \ref{prop1}. Thus by solving equation \eqref{proof1} we find the formal series $h$ and $T$
\begin{gather*}
h=\left(\begin{array}{cc}
1&0\\
0&v
\end{array}\right)+
\left(\begin{array}{cc}
-v_{1,0}&0\\
0&vv_{1,0}
\end{array}\right)\lambda^{-2}+
\left(\begin{array}{cc}
-vv_{1,0}&0\\
0&vv_{1,0}v_{2,0}+vv_{1,0}^2
\end{array}\right)\lambda^{-4}+\ldots,\\
T=\left(\begin{array}{cc}
1&0\\
0&1
\end{array}\right)+
\left(\begin{array}{cc}
0&0\\
1&0
\end{array}\right)\lambda^{-1}+
\left(\begin{array}{cc}
0&vv_{1,0}\\
v+v_{-1,0}&0
\end{array}\right)\lambda^{-3}+\nonumber\\
+\left(\begin{array}{cc}
0&vv_{1,0}v_{2,0}+vv_{1,0}^2\\
vv_{-1,0}+(v+v_{1,0})^2&0
\end{array}\right)\lambda^{-5}+\ldots.
\end{gather*}
The second operator of Lax pair diagonalized as follows
\begin{eqnarray}
&&S=D_m(T^{-1}\beta) g\beta^{-1}T=
\left(\begin{array}{cc}
1&0\\
0&\frac{u_{0,1}+\alpha_{0,1}}{u_{0,1}-\alpha_{0,1}}
\end{array}\right)\lambda+\nonumber\\
&&+\left(\begin{array}{cc}
2\alpha_{0,1}(u_{0,1}+\alpha_{0,1})&0\\
0&-\frac{2\alpha_{0,1}v}{u_{0,1}-\alpha_{0,1}}
\end{array}\right)\lambda^{-1}+
\left(\begin{array}{cc}
-\frac{2\alpha_{0,1}vv_{0,1}}{u_{0,1}-\alpha_{0,1}}&0\\
0&\frac{2\alpha_{0,1}vv_{1,0}}{u_{0,1}-\alpha_{0,1}}
\end{array}\right)\lambda^{-3}+\ldots.\nonumber
\end{eqnarray}

Comparing the coefficients before $\lambda$ in (\ref{logSH}) one gets the conservation laws. The first of them have a form
\begin{eqnarray}
&&(D_n-1)\ln\frac{u_{0,1}+\alpha_{0,1}}{u_{0,1}-\alpha_{0,1}}=(D_m-1)\ln v,\nonumber\\
&&(D_n-1)(2\alpha_{0,1}(u_{0,1}+\alpha_{0,1}))=(D_m-1)(-v),\nonumber\\
&&(D_n-1)\frac{-2\alpha_{0,1}v}{u_{0,1}+\alpha_{0,1}}=(D_m-1)v_{1,0},\nonumber\\
&&(D_n-1)\left(\frac{2\alpha_{0,1}vv_{1,0}}{u_{0,1}+\alpha_{0,1}}-\frac{2\alpha_{0,1}^2v^2}{(u_{0,1}+\alpha_{0,1})^2}\right)=
(D_m-1)\left(v_{1,0}v_{2,0}+\frac{1}{2}v_{1,0}^2\right).\nonumber
\end{eqnarray}

\section{Examples of semi-discrete models}

\subsection{Toda lattice}

As an example of application of diagonalization to differential-difference equations consider the well known Toda lattice
\begin{equation}\label{Toda}
\ddot{q}=e^{q_{1}-q}-e^{q_-q_{-1}},
\end{equation}
which is a compatibility condition of the following linear system
\begin{equation}\label{PsiToda}
\Psi_{1}=f\Psi_,\quad \Psi_{t}=g\Psi.
\end{equation}
Here the potentials are polynomials on the spectral parameter $\lambda$ of the form
\begin{equation}
f=\left(\begin{array}{cc}
p+\lambda&e^{q}\\
-e^{-{q}}&0
\end{array}\right),\quad 
g=\left(\begin{array}{cc}
0&-e^{q}\\
e^{-q_{-1}}&\lambda
\end{array}\right),
\end{equation}
where $p=\dot{q}$. In order to find the canonical form of the Lax operator we factorize the potential $f$ as $f=\alpha Z\gamma,$ where
\begin{gather*}
\alpha=\left(\begin{array}{cc}
1&0\\
-\frac{e^{-q}}{\lambda+p}&\frac{1}{1+p\lambda^{-1}}
\end{array}
\right),\quad 
Z=\left(\begin{array}{cc}
\lambda&0\\
0&\lambda^{-1}
\end{array}
\right),\quad 
\gamma=\left(\begin{array}{cc}
1+p\lambda^{-1}&e^{q}\lambda^{-1}\\
0&1
\end{array}
\right).
\end{gather*}
By the gauge transform $\phi=\gamma \Psi$ reduce the first equation in (\ref{PsiToda}) to the form $\phi_{10}=P(n,\lambda)Z\phi$ with
\begin{gather}
P(\lambda)=\gamma_{1}\alpha=\left(\begin{array}{cc}
1+p_{1}\lambda^{-1}-\frac{e^{q_{1}-q}\lambda^{-2}}{1+p\lambda^{-1}}&\frac{e^{q_{1}\lambda^{-1}}}{1+p\lambda^{-1}}\\
-\frac{e^{-q}\lambda^{-1}}{1+p\lambda^{-1}}&\frac{1}{1+p\lambda^{-1}}
\end{array}\right)=\\
=1+\left( \begin{array}{cc}
p_{1}&e^{q_{1}}\\
-e^{-q}&-p
\end{array}\right)\lambda^{-1}+
\left( \begin{array}{cc}
-e^{q_{1}-q}&-e^{q_{1}}p\\
e^{-q}p&p^2
\end{array}\right)\lambda^{-2}+
\left( \begin{array}{cc}
e^{q_{1}-q}p&e^{q_{1}}p^2\\
-e^{-q}p^2&-p^3
\end{array}\right)\lambda^{-3}+
\dots.
\end{gather}
Evidently $P(\lambda)$ is analytical at $\lambda=\infty$ and $det_jP(\infty)\neq 0$. By solving equation $D_nY\cdot h=P\cdot \bar Y$ one finds formal series $Y$ and $h$
\begin{gather*}
Y=\left( \begin{array}{cc}
1&0\\
0&1
\end{array}\right)+
\left( \begin{array}{cc}
0&0\\
-e^{-q_{-1}}&0
\end{array}\right)\lambda^{-1}+
\left( \begin{array}{cc}
0&0\\
e^{-q_{-1}}(p+p_{-1})&0
\end{array}\right)\lambda^{-2}+\\
\left( \begin{array}{cc}
0&-e^{q_{1}}\\
-e^{-q_{-1}}(p^2+pp_{-1}+p^2_{-1})-e^{-q_{-2}}-e^{q-2q_{-1}}&0
\end{array}\right)\lambda^{-3}+
\dots,\\
h=\left( \begin{array}{cc}
1&0\\
0&1
\end{array}\right)+
\left( \begin{array}{cc}
p_{1}&0\\
0&-p
\end{array}\right)\lambda^{-1}+
\left( \begin{array}{cc}
-e^{q_{1}-q}&0\\
0&p^2+e^{q_{1}-q}
\end{array}\right)\lambda^{-2}+\dots,\\
\gamma^{-1}=\left( \begin{array}{cc}
1&0\\
0&1
\end{array}\right)+
\left( \begin{array}{cc}
-p&e^{q}\\
0&0
\end{array}\right)\lambda^{-1}+
\left( \begin{array}{cc}
p^2&e^{q}p\\
0&0
\end{array}\right)\lambda^{-2}+
\left( \begin{array}{cc}
-p^3&-e^{q}p^2\\
0&0
\end{array}\right)\lambda^{-3}+\dots.
\end{gather*}
In order to diagonalize the second operator in (\ref{PsiToda}) we use the formal series $T=\gamma^{-1}Y$ where
\begin{gather*}
T=\left( \begin{array}{cc}
1&0\\
0&1
\end{array}\right)+
\left( \begin{array}{cc}
-p&-e^{q}\\
-e^{-q_{-1}}&0
\end{array}\right)\lambda^{-1}+
\left( \begin{array}{cc}
p^2+e^{q-q_{-1}}&e^{q}p\\
e^{-q_{-1}}(p+p_{-1})&0
\end{array}\right)\lambda^{-2}+\\
\left( \begin{array}{cc}
-p^3-e^{q-q_{-1}}(2p+p_{-1})&-e^{q}p^2-e^{q_{1}}\\
-e^{-q_{-2}}-e^{-q_{-1}}(p^2+pp_{-1}+p_{-1}^2)-e^{q-2q_{-1}}&0
\end{array}\right)\lambda^{-3}+\dots.
\end{gather*}
Then from (\ref{PsiToda}) we get
\begin{gather*}
S:=-T^{-1}T_t+T^{-1}gT=
\left( \begin{array}{cc}
0&0\\
0&1
\end{array}\right)\lambda-
\left( \begin{array}{cc}
1&0\\
0&1
\end{array}\right)+
\left( \begin{array}{cc}
e^{q_{1}-q}&0\\
0&-e^{q-q_{-1}}
\end{array}\right)\lambda^{-1}+\\
\left( \begin{array}{cc}
-e^{q_{1}-q}p&0\\
0&e^{q-q_{-1}}p
\end{array} \right) \lambda^{-2}
+\left( \begin{array}{cc}
e^{2q-2q_{-1}}+e^{q_{1}-q}p^2&0\\
0&s_{2,2}
\end{array}\right)\lambda^{-3}+\dots,\\
\mbox{where} \quad s_{2,2}=e^{q-q_{-2}}+e^{2q-2q_{-1}}+2e^{q_1-q_{-1}}+e^{q-q{-1}}(p^2_{-1}+2p^2+pp_{-1}).
\end{gather*}
Finally we give several conservation laws for the Toda lattice (see, also \cite{Toda})
\begin{eqnarray}
&&D_t p_{1}=(D_n-1)e^{q_{1}-q}, \nonumber \\ 
&&D_t\left(\frac{p_{1}^2}{2}+e^{q_{1}-q}\right)=(D_n-1)e^{q_{1}+q}p,\nonumber\\
&&D_t\left(\frac{p^2}{2}-e^{q_{1}-q}\right)=(D_n-1)(-e^{{q}-q_{-1}}p),\nonumber\\
&&D_t\left(-\frac{p^3}{3}-e^{q_{1}-q}(p_{1}+p)\right)=(D_n-1)(e^{2{q}-2q_{-1}}+e^{q_{1}-q}p^2).\nonumber
\end{eqnarray}

\subsection{Veselov-Shabat-Yamilov dressing chain}

Consecutive application of the B\"acklund transform to the Korteweg- de Vries equation generates a semi-discrete equation \cite{SY}, \cite{Shabat92},  \cite{ves}
\begin{equation}\label{VSeq}
f_{1,x}+f_{x}=f^2-f_1^2+\alpha-\alpha_1\quad \mbox{or} \quad 
u_{1,x}+u_x=(u_1-u)\sqrt{2(u_1+u)-4\alpha},
\end{equation}
\begin{equation}
u=f^2+f_x+\alpha,\quad u_1=f^2-f_x+\alpha
\end{equation}
called Veselov-Shabat-Yamilov dressing chain.
It admits the Lax pair
\begin{equation}\label{vslax}
\Psi_x=U\Psi, \quad \Psi_1=W\Psi,
\end{equation}
where 
\begin{equation}
U=\left( \begin{array}{cc}
0&1\\
u+\lambda^2&0
\end{array}\right), \quad
W=\left( \begin{array}{cc}
-f&1\\
f^2+\alpha+\lambda^2&-f
\end{array}\right).
\end{equation}
In order to diagonalize the Lax pair it is more convenient to begin with the first equation in (\ref{vslax}). Note that the leading term in $U$, i.e. the coefficient before $\lambda^2$ is a matrix with coinciding eigenvalues $\lambda_1=0$, $\lambda_2=0$. It makes impossible application of diagonalization method \cite{Dri} directly to the system. Let us first change the variables in such a way $\Psi =K\phi$, where 
\begin{equation}
K=\left( \begin{array}{cc}
1&1\\
\lambda&-\lambda
\end{array}\right)
\end{equation}
and reduce the system to the form
\begin{equation}\label{newvslax}
\phi_x=\bar U\phi, \quad \phi_1=\bar W\phi,
\end{equation}
with potentials
\begin{equation}\nonumber
\bar U=\sigma_3\lambda
+\frac{u}{2}R\lambda^{-1} \quad \mbox{and} \quad \bar W=\sigma_3\lambda-f+\frac{f^2+\alpha}{2}R\lambda^{-1}.
\end{equation}
\begin{equation}\nonumber
\sigma_3=\left(\begin{array}{cc}
1&0\\0&-1
\end{array}\right),\quad R=\left(\begin{array}{cc}
1&1\\-1&-1
\end{array}\right).
\end{equation}
Now the coefficient before $\lambda$ in $\bar U$ evidently has different eigenvalues $\lambda=\pm 1$. Therefore the method of formal diagonalization can be applied to the first equation in \eqref{newvslax}. Suppose that the formal series $T$ diagonalizes the operator $D_x-\bar U$ such that $D_x-h=T^{-1}(D_x-\bar U)T$. The unknowns 
\begin{equation}
T=1+T_1\lambda^{-1}+T_2\lambda^{-2}+\dots,\quad \mbox{and}\quad 
h=\sigma_3\lambda+h_0+h_1\lambda^{-1}+h_2\lambda^{-2}+\dots
\end{equation}
are found from the equation
\begin{gather}
\left(D_x-\sigma_3\lambda-\frac{u}{2}R\lambda^{-1}\right)\left(1+T_1\lambda^{-1}+T_2\lambda^{-2}+\dots\right)=\\
=\left(1+T_1\lambda^{-1}+T_2\lambda^{-2}+\dots\right)\left(D-\sigma_3\lambda-h_0-h_1\lambda^{-1}-h_2\lambda^{-2}-\dots\right).
\end{gather}
Omit the calculations and give only the answers
\begin{gather}
T=1+\lambda^{-1}+\left(\begin{array}{cc}
0&-\frac{u}{4}\\
-\frac{u}{4}&0
\end{array}\right)\lambda^{-2}+
\left(\begin{array}{cc}
0&-\frac{u}{4}-\frac{u_x}{8}\\
-\frac{u}{4}+\frac{u_x}{8}&0
\end{array}\right)\lambda^{-3}+\\
+\left(\begin{array}{cc}
0&\frac{u^2}{8}-\frac{u_x}{8}-\frac{u_{xx}}{16}\\
\frac{u^2}{8}+\frac{u_x}{8}-\frac{u_{xx}}{16}&0
\end{array}\right)\lambda^{-4}+\dots.
\end{gather}

\begin{gather}
h=\sigma_3\lambda+\left(\begin{array}{cc}
\frac{u}{2}&0\\
0&-\frac{u}{2}
\end{array}\right)\lambda^{-1}+
\left(\begin{array}{cc}
-\frac{u^2}{8}&0\\
0&\frac{u^2}{8}
\end{array}\right)\lambda^{-3}+
\left(\begin{array}{cc}
\frac{uu_x}{16}&0\\
0&\frac{uu_x}{16}
\end{array}\right)\lambda^{-4}+\dots.
\end{gather}

The second equation in (\ref{newvslax}) is reduced to the diagonal form $\psi_1=S\psi$ by the same transformation $\phi=T\psi$. The formal series  $S=T^{-1}\bar W T$ is of the form
\begin{gather}\nonumber
S=\sigma_3\lambda-fE+\frac{f^2+\alpha}{2}\sigma_3\lambda^{-1}-
\frac{(f^2+\alpha)u}{8}\sigma_3\lambda^{-3}+\frac{(f^2+\alpha)(f_{xx}+2ff_x)}{16}E\lambda^{-4}+\dots,
\end{gather}
where $E$ is the identity matrix. The diagonal series $h$ and $S$ produce an infinite set of conservation laws
\begin{equation}
D_x\ln S=(D_n-1)h.
\end{equation}
Let us give some of them in an explicit form
\begin{eqnarray}
D_xf&=&-(D_n-1)\frac{u}{2},\nonumber\\
D_x\left(\frac{f^3}{3}+\alpha f\right)&=&-(D_n-1)\frac{u^2}{4},\nonumber\\
D_xf_x(f^2+\alpha)&=&-(D_n-1)\frac{uu_x}{2}.\nonumber
\end{eqnarray}

\section{Evaluation of higher symmetries}

In this section we demonstrate how to use the diagonalization algorithm to evaluate hierarchies of semi-discrete generalized symmetries. We concentrate on a particular kind of discrete operators of canonical form
\begin{equation}\label{6operator}
L=D_n^{-1}cZ,
\end{equation}
where $Z=diag (\lambda, 1)$ is a diagonal matrix and $c=c(n,m)$ is an arbitrary function ranging on the Lie group GL(2,C) of $2\times 2$ non-degenerate matrices. Assume that the entries of the matrix $c=c(n,m)$ satisfy the following regularity  condition $c_{11}\neq0$. Due to the Proposition \ref{prop1} operator $L$ under this condition can be diagonalized. This means that there exist two formal series $T=\sum_{k=0}^{\infty}T(k)\lambda^{-k}$ and $h=\sum_{k=0}^{\infty}h(k)\lambda^{-k}$ such that $L_0:=T^{-1}LT$ is an operator with diagonal coefficients $L_0=D_n^{-1}hZ$. This representation allows one to construct semi-discrete models for which $L$ serves as the Lax operator. Following the procedure developed in reference \cite{Dri} we describe the class $Z_L$ of the formal series of the form 
\begin{equation}\label{6A}
A=\sum_{k=-N}^{\infty}A(k)\lambda^{-k}
\end{equation}
commuting with $L$. Here $A(k)$ are $2\times 2$ matrices. By applying the conjugation to the equation $[L,A]=0$ one gets
\begin{equation}\label{6A0}
[L_0,A_0]=0, \quad \mbox{where} \quad A_0:=T^{-1}AT=\sum_{k=-N}^{\infty}A_0(k)\lambda^{-k}
\end{equation}

\begin{lem}\label{lem1} Coefficients $A_0(k)$ of the series (\ref{6A0}) are diagonal matrices which do not depend on $n$. 
\end{lem}
{ \bf Scheme of proof}. The commutativity relation $[L_0,A_0]=0$ implies $D_n(A_0)h =hZA_0Z^{-1}$. This equation looks similar to (\ref{proof1}) where the potential $P(n,\lambda)$ in the right side is replaced by a diagonal factor $h$. Comparing the coefficients of powers of $\lambda$ one can easily complete the proof of the Lemma.

The statement converse to that of Lemma is also true. For any formal series $A_0=\sum_{k=-N}^{\infty}A_0(k)\lambda^{-k}$ with diagonal coefficients $A_0(k)$ which do not depend on $n$ the series $A=TA_0T^{-1}$ is in $Z_L$. 

{\bf Remark}. It is important that for any $k$ the coefficient $A(k)$ depends on a finite number of shifts of the variable $c$. 

Thus we have a complete description of the set $Z_L$.

Now by using $Z_L$ one can construct a hierarchy of commuting semi-discrete models associated with the operator $L$. Let us take an arbitrary $A\in Z_L$ and split it down as a sum of two parts $A=A_++A_-$, where $A_+=\sum_{k=-N}^{-1}A(k)\lambda^{-k}+B_+$ and $A_-=\sum_{k=1}^{\infty}A(k)\lambda^{-k}+B_-$. Matrices $B_{\pm}$ are chosen as follows
\begin{equation}\label{BB}
B_+=\left(\begin{array}{cc}
*&*\\
0&*
\end{array}\right),\quad 
B_-=\left(\begin{array}{cc}
*&0\\
*&*
\end{array}\right).
\end{equation}
There is a freedom in choosing of diagonal entries of the matrices $B_+$, $B_-$. Choose them as functions depending on a finite number of the matrices $A(k)$. For instance one can take  
\begin{equation}\label{BB00}
B_+=\left(\begin{array}{cc}
*&*\\
0&*
\end{array}\right),\quad 
B_-=\left(\begin{array}{cc}
0&0\\
*&0
\end{array}\right).
\end{equation}
In virtue of the conditions (\ref{BB}) we have that both functions $A_+$ and $ZA_+Z^{-1}$ are polynomials in $\lambda$ while the series $A_-$ and $ZA_-Z^{-1}$ do not contain terms with positive powers of $\lambda$.
\begin{pro}\label{prop2}
The consistency condition of the following system of linear equations 
\begin{equation}\label{LA6}
L\phi=\phi, \quad \phi_t=A_+\phi
\end{equation}
is equivalent to a differential-difference equation of the form
\begin{equation}\label{eq6}
\frac{d}{dt}c=F(c,c_{\pm1},c_{\pm2},...c_{\pm r}).
\end{equation}
\end{pro}
{\bf Proof}. Substitute the representation $A=A_++A_-$ into the commutativity relation $[L,A]=0$ and find
\begin{equation}\label{r6}
r(\lambda):=cZA_+Z^{-1}-D_n(A_+)c=-cZA_-Z^{-1}+D_n(A_-)c.
\end{equation}
Due to the condition (\ref{BB}) we have  $r(\lambda)\equiv const$. Indeed, the right hand side of (\ref{r6}) is a polynomial, but the left side does not contain positive powers of $\lambda$. Therefore by taking $r=\frac{d}{dt}c$ one gets a self-consistent equation. Evidently equation (\ref{r6}) generates the relations $[L,\frac{d}{dt}-A_+]=0$ and $[L,\frac{d}{dt}-A_-]=0$ the former of which is nothing else but the consistency condition of the system (\ref{LA6}). Note that due to the Remark above and the choice of the normalization (\ref{BB}) the coefficients of the polynomial $A_+=A_+(\lambda)$ are functions of a finite number of the variables $c$, $c_{\pm1}$, $c_{\pm2}, ...$.

System (\ref{eq6}) has too much freedom. Actually it is preserved under the following type change of variables $c'=D_n(w^{-1})cw$, where $w$ satisfies the condition $[w,Z]=0$. This change of variables generates a gauge transform of the system (\ref{LA6}) as $L\rightarrow w^{-1}Lw$ and $A\rightarrow w^{-1}Aw-w^{-1}w_t$. In order to restrict this freedom one has to impose some additional constraint. Constraint 
\begin{equation}\label{constraint}
c_{11}=const,  \quad c_{22}=const,
\end{equation}
reduces operator (\ref{6operator}) to the Lax operator of the Ablowitz-Ladik hierarchy \cite{Abl-Lad}.
The constraint is consistent with the dynamics (\ref{eq6}) if the normalization is chosen as (\ref{BB00}). Indeed, comparison of the coefficients before $\lambda^0$ in the equation $\frac{d}{dt}c=-cZA_-Z^{-1}+D_n(A_-)c$ gives  $\frac{d}{dt}c_{11}=\frac{d}{dt}c_{22}=0$. 

Note that there is another diagonalization of the operator (\ref{6operator}). Indeed, if $c_{22}\neq0$, $\det c\neq0$ then due to the second part of the Proposition \ref{prop1} one can construct formal series with positive powers of $\lambda$, i.e. $T'=\sum_{k=0}^{\infty}T'(k)\lambda^{-k}$ and $h'=\sum_{k=0}^{\infty}h'(k)\lambda^{-k}$ such that the operator $L'_0:=T'^{-1}LT'$ is of the form $L'_0=D_n^{-1}h'Z$ where the coefficients $h'(k)$ of the formal series $h'$ are diagonal matrices, moreover the coefficients $T'(k)$ of the series $T'$ depend on a finite number of the variables $c, D^{\pm1}_nc, D^{\pm2}_nc, ...$. Due to the reasonings above this diagonalization allows one to construct another hierarchy of symmetries. Suppose that $c_{11}=c_{22}=1$, $c_{21}=u$, $c_{12}=v$ then the members of  both hierarchies are symmetries of the following system of discrete equations
\begin{eqnarray}\label{dissystem}
u_{1,0}-u_{0,1}&=&\varepsilon u_{1,1}(1-u_{0,1}v_{1,0})\\
v_{0,1}-v_{1,0}&=&\varepsilon v(1-u_{0,1}v_{1,0})\nonumber
\end{eqnarray}
admitting the Lax pair of the form
\begin{equation}
L=D^{-1}_n\left(\begin{array}{cc}
1&v\\
u&1
\end{array}\right)Z,\quad 
M=D^{-1}_m\left(\begin{array}{cc}
\lambda+\varepsilon (1-u_{-1,1}v)&v\\
\lambda u_{-1,1}&1
\end{array}\right).
\end{equation}

One can prove by evaluating the continuum limit that \eqref{dissystem} is the lattice version of the NLS equation (see \S 7 below). Let us illlustrate the method of evaluating the higher symmetries with the following example. 

{\bf Example}. By taking $c_{11}=1$, $\det c=1$ one reduces the coefficient of the operator (\ref{6operator}) to the form (see \eqref{fgidlns})
$$c=\left(\begin{array}{cc}
1&v\\
u&1+uv
\end{array}\right).$$
To find the diagonalization of the operator we use Proposition \ref{prop1}. Choose the starting matrix $A_0=diag(\lambda/2,-\lambda/2)$ with vanishing trace and define $A=T^{-1}A_0T$ commuting with $L$, where $T$ is given by (\ref{Tseries}). Let us specify first two terms in the series $A=A_{-1}\lambda+A_0+A_1\lambda^{-1}+\dots$ 
$$A=\left(\begin{array}{cc}
1/2&0\\
u_{-1}&-1/2
\end{array}\right)\lambda+\left(\begin{array}{cc}
-u_{-1}v&v\\
u_{-2}-u_{-1}^2v&u_{-1}v
\end{array}\right)+\dots.$$
Due to the normalization (\ref{BB00}) we take 
$$A_+=\left(\begin{array}{cc}
1/2&0\\
u_{-1}&-1/2
\end{array}\right)\lambda+\left(\begin{array}{cc}
-u_{-1}v&v\\
0&u_{-1}v
\end{array}\right).$$
Then the commutativity relation $[L,\frac{d}{dt}-A_+]=0$ leads to the well-known system of  equations \cite{SY}
\begin{eqnarray}
&&u_t=-u_{-1}+uv^2,\\
&&v_t=v_1-u^2v
\end{eqnarray}

To conclude this section, we present one more system of equations connected with the further reduction of the operator (\ref{6operator})
\begin{eqnarray}\label{dissystem2}
a_{1,0}(1-\kappa u^2)&=&a(1-\kappa u^2_{0,1}),\quad \kappa=\pm 1,\\
a_{1,0}(u-u_{1,1})&=& u_{0,1}-u_{1,0})\nonumber
\end{eqnarray}
which is the commutativity condition of the operators
\begin{equation}
L=D^{-1}_n\left(\begin{array}{cc}
1&\kappa u\\
u&1
\end{array}\right)Z,\quad 
M=D^{-1}_m\left(\begin{array}{cc}
\lambda+a&\kappa(u-au_{0,1})\\
\lambda (u-au_{0,1})&a\lambda+1
\end{array}\right).
\end{equation}
Excluding the variable $a$ from the system (\ref{dissystem2}) one obtains a six-points discrete equation 
\begin{equation}\label{dissystem4}
\frac{(u_{0,1}-u_{1,0})(1-ku^2)}{u-u_{1,1}}=\frac{(u_{-1,1}-u)(1-ku^2_{0,1})}{u_{-1,0}-u_{0,1}}.
\end{equation}
By evaluating the continuum limit of the equation (\ref{dissystem4}) one gets the following equation
\begin{equation}
kuu_x(u_x^2-u_t^2)+2(1-ku^2)(u_tu_{xt}-u_xu_{tt})=0.
\end{equation}

\section{Examples of matrix models}
\subsection{Lattice version of the ``matrix'' NLS}
System (\ref{dissystem}) admits a simple matrix generalization
\begin{eqnarray}\label{dissystem3}
U_{1,0}-U_{0,1}&=&\varepsilon U_{1,1}(E_1-V_{1,0}U_{0,1}),\\
V_{0,1}-V_{1,0}&=&\varepsilon (E_1-V_{1,0}U_{0,1})V.\nonumber
\end{eqnarray}
Its Lax operator is given in terms of the block matrices 
\begin{equation}
L=D^{-1}_n\left(\begin{array}{cc}
\lambda E_1&V\\
\lambda U&E_2
\end{array}\right),\quad 
M=D^{-1}_m\left(\begin{array}{cc}
(\lambda+\varepsilon)E_1-\varepsilon V U_{-1,1}&V\\
\lambda U_{-1,1}&E_2
\end{array}\right).\label{66LA}
\end{equation}
Here $E_1$, $E_2$ are the identity matrices of the size $k$ and $l$, the field variables $U$ and $V$ are $l\times k$ and $k\times l$ matrices respectively.

Operator $L$ in (\ref{66LA}) is of canonical form since it can be represented as $L=D^{-1}_n cZ$ but it does not satisfy the requirements of the Proposition \ref{prop1}. Here we have $Z=diag (\lambda,...\lambda,1...1)$ and some of the exponents $\gamma_i$ coincide. However, the operator can be ``diagonalized'' in an appropriate sense. In this case the formal series $T$, $h$ and $S$ are block matrices
\begin{equation}
T(\lambda)=1+\sum_{k=0}^{\infty}\lambda^{-k}\left(\begin{array}{cc}
0&T_{12}(k)\\
T_{21}(k)&0
\end{array}\right),\quad 
h(\lambda)=\sum_{k=0}^{\infty}\lambda^{-k}\left(\begin{array}{cc}
h_{11}(k)&0\\
0&h_{22}(k)
\end{array}\right).\nonumber
\end{equation}
\begin{equation}
S(\lambda)=\sum_{k=l}^{\infty}\lambda^{-k}\left(\begin{array}{cc}
S_{11}(k)&0\\
0&S_{22}(k)
\end{array}\right).\nonumber
\end{equation}
As a result we get 

\begin{eqnarray}\nonumber
&&T=\left( \begin{array}{cc}
E_1&0\\
(E_2-UV)U_{-1,0}&E_2
\end{array}\right)+
\left( \begin{array}{cc}
0&0\\
*&0
\end{array}\right)\lambda^{-1}+\dots,\\\nonumber
&&h=\left( \begin{array}{cc}
E_1&0\\
0&E_2-U_{1,0}V_{1,0}
\end{array}\right)+
\left( \begin{array}{cc}
V_{1,0}U&0\\
0&-UV_{1,0}+U_{1,0}V_{1,0}UV_{1,0}
\end{array}\right)\lambda^{-1}+\dots,\\\nonumber
&&S=\left( \begin{array}{cc}
E_1&0\\
0&0
\end{array}\right)\lambda+
\left( \begin{array}{cc}
\varepsilon E_1-\varepsilon VU_{-1,1}+V_{0,1}U_{-1,1}&0\\
0&*
\end{array}\right)+\dots.
\end{eqnarray}

Conservation laws are deduced from the equations
\begin{equation}\label{7cl}
(D_m-1)\log\det h_{jj}=(D_n-1)\log\det S_{jj}, \quad j=1,2.
\end{equation}

Choose the matrices $U$ and $V$ in (\ref{66LA}) as a column and a row such that
\begin{equation}\label{UTVrow}
U^T=(u^{(1)}, u^{(2)},...,u^{(k)}), \quad V=(v^{(1)}, v^{(2)},...,v^{(k)})
\end{equation}
then evidently the product $V_{1,0}U_{0,1}=\sum^k_{j=1}u_{0,1}^{(j)}v_{1,0}^{(j)}$ is a scalar. For this case system (\ref{dissystem3}) is written as follows
\begin{eqnarray}\label{dvnls}
&&u^{(i)}_{1,0}-u^{(i)}_{0,1}=\left(1-\sum_{j=1}^{k}u_{0,1}^{(j)}v_{1,0}^{(j)}\right)\varepsilon u^{(i)}_{1,1},\\
&&v^{(i)}_{0,1}-v^{(i)}_{1,0}=\left(1-\sum_{j=1}^{k}u_{0,1}^{(j)}v_{1,0}^{(j)}\right)\varepsilon v^{(i)}\nonumber
\end{eqnarray}
and can be interpreted as integrable discretization of the ``vector" nonlinear Schr\"odinger equation describing the transmitting of the polarized pulses along optical fibres (see \cite{man}). The formal continuum limit in the lattice \eqref{dvnls} is evaluated below. System (\ref{dvnls}) admits an infinite sequence of conservation laws. They can be deduced from the formula (\ref{7cl}). The first one is of the form
\begin{eqnarray}
&&(D_n-1)\ln \frac{(1-VU_{-1,1})(1-V_{0,1}U_{0,1})}{1-VU}=(D_m-1)\ln (1-V_{1,0}U_{1,0}),\nonumber\\
&&(D_n-1)\left(V_{0,1}U_{-1,1}-\varepsilon VU_{-1,1}\right)=(D_m-1)\left(V_{1,0}U\right).
\end{eqnarray}

\subsection{Lattice version of the ``vector'' derivative NLS equation}
System \eqref{lnls} also admits a ``vector'' generalization:
\begin{eqnarray}\label{matrixLNLS}
&&U_{0,1}-U_{1,0}+U_{1,1}(V_{1,0}(U_{0,1}-U_{1,0})+\varepsilon)=0,\\
&&V_{1,0}-V_{0,1}+((V_{1,0}-V_{0,1})U_{0,1}+\varepsilon)V=0\nonumber
\end{eqnarray}
where $U$ and $V$ are column and respectively row vectors, defined in \eqref{UTVrow}. The system \eqref{matrixLNLS} is the commutavity condition of the  operators 
\begin{equation}\label{pairMlnls}
L=D_n^{-1}\left( \begin{array}{cc}
\lambda +VU&V\\
U&E
\end{array}\right),\quad
M=D_m^{-1}\left( \begin{array}{cc}
(\lambda+\varepsilon)+VU_{-1,1}&V\\
U_{-1,1}&E
\end{array}\right)
\end{equation}
where $E$ is the identity matrix of the size $k-1$. In the coordinate representation system \eqref{matrixLNLS} takes the form

\begin{eqnarray}
u_{0,1}^{(i)}-u^{(i)}_{1,0}+u_{1,1}^{(i)}\left(\sum_{j=1}^k (v_{1,0}^{(j)}u^{(j)}_{0,1}-v_{1,0}^{(j)}u^{(j)}_{1,0})+\varepsilon\right)=0,\nonumber\\
v_{1,0}^{(i)}-v^{(i)}_{0,1}+v^{(i)}\left(\sum_{j=1}^k (v_{1,0}^{(j)}u^{(j)}_{0,1}-v_{0,1}^{(j)}u^{(j)}_{0,1})+\varepsilon\right)=0.\nonumber
\end{eqnarray}
In the next section we prove that (\ref{matrixLNLS}) is the lattice version of the ``vector'' derivative NLS equation.

\subsection{Evaluation of the continuum limits}
In this section we show by evaluating formal continuum limits that the systems \eqref{dissystem3} and \eqref{matrixLNLS} are discretizations of the matrix NLS and respectively ``vector'' derivative NLS equations. Let us begin with \eqref{dissystem3}. Set $U(n,m)=r(t,x)$, $V(n,m)=s(t,x),$ $x=m\delta$, $t=\frac{\delta^2}{2}n$, $\varepsilon=\frac{\delta^2}{2}$. Then evidently the shifted variables are 
\begin{eqnarray}
&&U_{-1,0}=r-\frac{\delta^2}{2}r_t+o(\delta^2), \delta\mapsto 0,\nonumber\\
&&V_{1,0}=s+\frac{\delta^2}{2}s_t+o(\delta^2), \delta\mapsto 0,\\
&&U_{0,-1}=r-\delta r_x+\frac{\delta^2}{2}r_{xx}+o(\delta^2), \delta\mapsto 0,\nonumber\\
&&V_{0,1}=s+\delta s_x+\frac{\delta^2}{2}s_{xx}+o(\delta^2), \delta\mapsto 0.\nonumber
\end{eqnarray}
 
Now substitute the expressions found into the system \eqref{dissystem3} rewritten as follows
\begin{eqnarray}
&&U_{0,-1}-U_{-1,0}=U(E_1-V_{0,-1}U_{-1,0}),\nonumber\\
&&V_{0,1}-V_{1,0}=\varepsilon(E_1-V_{1,0}U_{0,1})V.\nonumber
\end{eqnarray}
After some simplifications one gets 
\begin{eqnarray}
&&r_t-\frac{2}{\delta}r_x+r_{xx}=r(E_1-sr)+o(\delta),\nonumber\\
&&-s_t+\frac{2}{\delta}s_x+s_{xx}=(E_1-sr)s+o(\delta), \quad \delta\mapsto 0.\nonumber
\end{eqnarray}
Introduce new variables by means of the equations $x=x'-\frac{2}{\delta}t'$, $t=t'$, $r=e^tR$, $s=e^{-t'}S$ then due to the relations $\frac{\partial}{\partial t'}=\frac{\partial}{\partial t}-\frac{2}{\delta}\frac{\partial}{\partial x}$, $\frac{\partial}{\partial x'}=\frac{\partial}{\partial x}$ one obtains for $\delta\mapsto 0$ the matrix NLS equation:
\begin{eqnarray}
&&R_{t'}+R_{x'x'}+RSR=0,\nonumber\\
&&-S_{t'}+S_{x'x'}+SRS=0.\nonumber
\end{eqnarray}

Show that system \eqref{matrixLNLS} is the lattice version of the ``vector'' derivative NLS equation. To this end evaluate the continuum limit of the system \eqref{matrixLNLS} by setting $U(n,m)=\delta r(x,t)$, $V(n,m)=\delta s(x,t)$, $x=\delta^2m$, $t=\frac{\delta^4}{2}n$, $\varepsilon=\delta^5$. Find now the Taylor expansions for the shifted variables
\begin{eqnarray}
&&U_{-1,0}=\delta r-\frac{\delta^5}{2}r_t+o(\delta^5), \delta\mapsto 0,\nonumber\\
&&V_{1,0}=\delta s+\frac{\delta^5}{2}s_t+o(\delta^5), \delta\mapsto 0,\\
&&U_{0,-1}=\delta r-\delta^3 r_x+\frac{\delta^5}{2}r_{xx}+o(\delta^5), \delta\mapsto 0,\nonumber\\
&&V_{0,1}=\delta s+\delta^3 s_x+\frac{\delta^5}{2}s_{xx}+o(\delta^5), \delta\mapsto 0.\nonumber
\end{eqnarray}

Then substitute the values found into the system \eqref{matrixLNLS} rewritten as follows
\begin{eqnarray}
&&U_{-1,0}-U_{0,-1}+U(V_{0,-1}(U_{-1,0}-U_{0,-1})+\varepsilon)=0,\nonumber\\
&&V_{1,0}-V_{0,1}+((V_{1,0}-V_{0,1})U_{0,1}+\varepsilon)V=0.\nonumber
\end{eqnarray}
As a result one gets 
\begin{eqnarray}\label{rsNLS}
-\frac{\delta^5}{2}r_t+\delta^3r_x-\frac{\delta^5}{2}r_{xx}-\delta^5rsr_x=o(\delta^5),\\
\frac{\delta^5}{2}s_t-\delta^3s_x-\frac{\delta^5}{2}s_{xx}-\delta^5s_xrs=o(\delta^5).\nonumber
\end{eqnarray}
Introduce the new independent variables $t', x'$ by taking $x=x'-\frac{2}{\delta^2}t'$, $t=t'$ and due to the relations $\frac{\partial}{\partial x'}=\frac{\partial}{\partial x}$, $\frac{\partial}{\partial t'}=\frac{\partial}{\partial t}-\frac{2}{\delta^2}\frac{\partial}{\partial x}$ find the limit for $\delta\mapsto 0$. Then \eqref{rsNLS} yields the ``vector''  version of the derivative NLS equation
\begin{eqnarray}
r_{t'}+\frac{1}{2}r_{x'x'}+rsr_{x'}=0,\nonumber\\
-s_{t'}+\frac{1}{2}s_{x'x'}+s_{x'}rs=0.\nonumber
\end{eqnarray}

\section{Conclusions}

Method of formal diagonalization suggested in \cite{Dri} is adopted to the case of discrete Lax operators. It is shown that diagonalization allows one to construct infinite sequences of conservation laws for discrete and semi-discrete dynamical systems. Efficiency of the method is illustrated with an important and representative set of examples, including the well known models like dpkdv, Toda lattice, dressing chain, Ablowitz-Ladik hierarchy as well as recently introduced models like Toda field equation with discrete space-time corresponding to the Lie algebra $A_1^{(1)}$. For a special class of discrete operators a hierarchy of higher symmetries is described via formal diagonalization. An example of the six-point integrable discrete models is presented. Systems of quad graph equations are found including lattice versions of the ''matrix NLS and ``vector'' derivative NLS equations. 

The article has some intersections with the work by A.V. Mikhailov \cite{Mikh2012}.

\section*{Acknowledgments}

This work is partially supported by Russian Foundation for Basic Research (RFBR) grants
11-01-97005-r-povoljie-a, 12-01-31208-mol\_a and 13-01-00070-a and by Federal Task Program
``Scientific and pedagogical staff of innovative Russia for 2009--2013''  contract no. 2012-1.5-12-
000-1003-011.

\end{document}